\documentclass[12pt]{amsart}
\usepackage{geometry,multicol}                 
\usepackage{graphicx} 
\usepackage{amssymb} 
\usepackage{epstopdf} 
\def\dj{d\kern-0.4em\char"16\kern-0.1em}

\title{Umbral Dynamics in the Near Infrared Continuum}
\author{A.~An\dj {\,}i\'c$^1$, W.~Cao$^1$, P.~R.~Goode$^1$}

\begin{document}
\maketitle
{\small1. Big Bear Solar Observatory, 40398 North Shore Lane, Big Bear City, CA-92314, USA.}

{\tiny.\\
Submitted to ApJ: 10 December 2010\\
Accepted in ApJ: 4 May 2011}

\section{Abstract}

{\it We detected peaks of oscillatory power at $3$ and $\sim 6.5$ minutes in the umbra of the central sunspot of the active region NOAA AR 10707 in  data obtained in the near infrared (NIR) continuum at $1565.7$ nm. The NIR dataset captured umbral dynamics around $50$ km below the $\tau_{500}=1$ level. The umbra does not oscillate as a whole, but rather in distinct parts that are distributed over the umbral surface. The most powerful oscillations, close to a period of $\sim 6.5$, do not propagate upward. We noted a plethora of large umbral dots that persisted for $\geq 30$ minutes and stayed in the same locations. The peaks of oscillatory power above the detected umbral dots are located at $3$ and $5$ minutes oscillations, but are very weak in comparison with the oscillations of $\sim 6.5$ minutes.}

\vskip11mm

\begin{multicols}{2}

\section{Introduction}

The umbra, the darkest part of a sunspot, has complex dynamics. The umbral oscillations,  part of the umbral dynamics,  have been the subject of many research studies. At chromospheric levels, researchers have observed velocity oscillations with periods close to $3$ minutes and velocity amplitudes of about $6$ km/s (Balthasar and Wiehr, 1984; Soltau and Wiehr, 1984; Tirospula et al., 2000). At the photospheric level, umbral oscillations have periods closer to $5$ minutes, with velocity amplitudes of about $75$ m/s (Lites, 1992).\par
 From a theoretical standpoint, there are two postulated mechanisms for the generation of umbral oscillations. One is that they are generated on sub-photospheric levels by a common extended source (Rouppe van der Voort et al. 2003; Bogdan and Judge, 2006) and propagate upward along the magnetic field lines (L\'{o}pez Aristre et al., 2001). The other postulate suggests that umbral oscillations are generated when slow mode waves are trapped  within a cavity in the umbra's chromosphere. Broadband acoustic noise from the convection zone excites the trapped waves (Zhugzhda et al.,1983). Lites (1986) found that the $5$ minute photospheric oscillations do not drive the $3$ minute chromospheric umbral  oscillations.\par 
 The umbra does not oscillate as the whole, but in distinct areas within the umbra (Socas-Navarro et al. 2009). Socas-Navarro et al. (2009) found that the umbra is tremendously dynamic and requires a time cadence faster than 20 s to resolve the apparent motions of the emission source.  \par  
 Umbral dots (UDs), small-scale structures distributed across the umbra, are part of umbral dynamics. Sch\"{u}ssler and V\"{o}gler, (2006) showed that the appearance of UDs is a natural consequence of magnetoconvection under the influence of a strong magnetic field. This was confirmed by Cheung et al. (2010) and Bharti et al. (2010). Heinemann et al. (2007) stated that UDs are caused by overturning convection. \par
Balthasar and Wiehr (1984) stated that the lifetime of UDs is on the order of $20$ minutes. Rimmele (2008), Riethm\"{u}ller et al. (2008), and Ortiz et al. (2010) observed the signature of UDs dynamics, as described by the model of Sch\"{u}ssler and V\"{o}gler (2006). Rimmele (2008) noted that UDs have a lifetime close to 30 minutes, as predicted  by the model of Sch\"{u}ssler and V\"{o}gler (2006). Bharti et al. (2010) similarly predicted that the average UD's lifetime is between $25$ and $28$ minutes. On the other hand, Hamedivafa (2008) stated that the average lifetime of the UDs is between $7$ and $10$ minutes, while Watanabe et al. (2009) measured the average lifetime as $7.3$ minutes. \par 
Socas-Navarro et al. (2009) detected a movement of the bright umbral elements. They registered horizontal propagation speeds of 30 km/s and stated that their cadence of 20 s is not enough to resolve the fast lateral motion of the oscillatory sources in the umbra. \par
In this work, we present observations of UDs and umbral oscillations at a level of $50$ km below the $\tau_{500}=1$. The NIR continuum provides easier seeing correction with adaptive optics (AO), and much lower scattered light, both instrumental and atmospheric. Hence we were able to observe structures near the telescope's diffraction limit for this wavelength even without image reconstruction. 

\section{Observations and Data analysis}

High-resolution photometric observations of the solar active region NOAA AR10707 were obtained simultaneously in G-band and the near infrared (NIR) continuum near $1.6 \mu$m with the Dunn Solar Telescope (DST) at the National Solar Observatory/Sacramento Peak (NSO/SP) on December 1, 2004. Benefiting from a high order AO system, the spatial resolution was close to the diffraction limit of the $76$ cm aperture DST in the NIR continuum, but significantly poorer for G-band. \par 

To probe the deepest layer of the solar atmosphere, exploiting the Sun's opacity minimum at $1.6 \mu$m, NIR photometry was employed, which consists of a tuneable NIR birefringent Lyot filter developed by the Center for Solar-Terrestrial Research/NJIT (Cao et al. 2006), an interference filter with a $5$ nm bandpass, and a NIR camera. The Lyot filter has a very narrow bandpass of $0.25$ nm and was tuned to the line-free continuum at $1565.7$ nm. The NIR camera (Cao et al. 2005) is based on a $1024 \times 1024$ HgCdTe/Al2O3 CMOS focal plane array with a liquid nitrogen cooling system. The pixel size is $18 \mu$m$ \times 18 \mu$m. The output signal is digitalised into 14-bits with a dynamic range better than $70$ dB. \par 
The field of view (FOV) was $122'' \times 122''$ and $105'' \times 105''$ for the NIR and the G-band, respectively. Ten NIR frames were obtained every second, and the best frame was selected. The G-band data set had a cadence of $0.5$ s. All images were dark- and flat-field corrected and selected according to the highest root mean square (RMS) contrast. The observational run started on 16:54 UT and produced datasets $30$ minutes long.\par 
To equalise the time cadence of both datasets, we took every second image from the G-band dataset, achieving a $1$ s cadence. Thus in both dataset we acquired a Nyquist period of $2$ s. \par 

The data were co-aligned using a Fourier coaligning routine; which uses cross-correlation techniques and squared mean absolute deviations to provide sub-pixel coalignment precision. However, we did not implement sub-pixel image shifting, to avoid substantial interpolation errors that sometimes accompany the use of this technique. The reference image for coaligning routine was the floating mean; we made a mean image for each interval of 30 data images, shifting the interval by one image over the time series.
De-stretching of the images was performed to eliminate the influence of seeing distortions. The de-stretching routine uses bilinear interpolation. The reference image for the de-stretching routine is the floating mean; we made a mean image for each interval of 10 data images, shifting the interval by one image over the time series.\par 

Wavelet analysis based on work of  Torrence and Compo (1998) was applied.  The wavelet analysis most useful output is the power of the detected oscillatory signal that is related to the amplitude of the analysed oscillatory signal. (Vigouroux and Delache, 1993; Graps, 1995; Starch et al. 1997). In our analysis we used power of the oscillations as the main product of the wavelet analysis. The oscillatory detection was performed with use of the Morlet wavelet as the mother wavelet:

\begin{equation}
\psi_0(t)= \pi^{-\frac{1}{4}}e^{i \omega_0 t} e^{-\frac{t^2}{2}},
\label{vaveleti1}
\end{equation}

\noindent where $\omega_0$ is the non-dimensional frequency and $t$ the non-dimensional time parameter. Morlet wavelet is non-orthogonal, based on the Gaussian function and  very close to the limit of the signal processing uncertainty, ($\sqrt{\pi}$). An associated Fourier period, P, corresponds to $1.03s$, where $s$ is the wavelet scale.  The wavelet transform is a convolution of the time series with the analysing wavelet function.  The complete wavelet transform is achieved by varying the wavelet scale, which controls both the period and temporal extent of the function. Only one-dimensional wavelet analysis is done for each spatial coordinate. At the beginning and end of the wavelet transforms are regions where spurious power may arise as a direct result of the finite extent of the time series. Those regions are usually refereed to as the cone of influence (COI), having a temporal extent equal to the $e$-folding time of the wavelet function. For the case of our wavelet is:

\begin{equation}
t_d= \sqrt{2s}=\sqrt{2}\frac{P}{1.03},
\label{vaveleti2}
\end{equation}

This time scale is the response of the wavelet function to noise spikes and is used in our detection criteria by requiring that oscillations have a duration greater than $t_d$. The finite nature of our time series compelled us to impose the maximum period of $12$ minutes above which we did not accept any detection. The approach to automated wavelet analysis used here and all the imposed restrictions are detailed in Andic et al. (2010).
Since the data used contained seeing distortions, we took into considerations only the oscillations that had 99\% confidence levels.  \par 
The diameter of the UDs was measured as the full-width at half maximum (FWHM) of a spatial profile that contained the maximum intensity of the structure. This analysis was performed at the frame with the best contrast. Determination of the lifetime and the UDs shift in the horizontal plane was performed with  the nonlinear affine velocity estimator (NAVE) method (Chae and Sakurai, 2008), which tracked the detected structures through the dataset.  \par 
The areas surrounding the umbra were masked out with a binary mask. We made an average frame over the time series and normalised it to the maximum. Then, we multiplied all  pixels with brightness $\geq 0.75$ by $0$ and darker pixels by $1$, masking out  in such way everything but umbra. \par
For the phase analysis of the oscillations two methods were applied. One is wavelet phase analysis between NIR and G-band data sets. This phase analysis is similar to the Fourier phase analysis, only it provides  time and frequency localisation.  This localisation is provided by  the use of wave packets by wavelet analysis instead of the infinite wave train of  Fourier analysis.   The difference in cyclic phase, $\Delta\phi$ can be determined for each frequency component, $\nu$, using the phase information contained within the complex wavelet transform. The quality of the values is represented by phase coherence. A time series in this work are extracted from the same ($x$,$y$) pixel location in both datasets. The signals are separated in the direction normal to the solar surface. Therefore, the phase differences can be interpreted as delays caused by the finite propagation speed of waves traveling between the optical formation heights. The automated method used is described in detail in Bloomfield et al. (2004). \par
The second method was combining the Fourier and Hilbert transform on a single dataset (White and Cha, 1973). In short, the signal is a complex function. The real part is original signal and imaginary part is the quadrature of the original signal. Since a real function and its quadrature are Hilbert transform pairs, the Hilbert transform converts one into another.  Resulting transform describes the amplitude and phase of a variable in complex plane. The signal is transformed into the Fourier space then transformed back using the Hilbert transform (Stebbins and Goode, 1987). \par
The phase angle spectrum is formed in such way that an upward propagating wave has a positive phase angle. The phase angles are presented in a weighted diagram, where weighting is applied per sample by cross-power amplitude $\sqrt{P_1P_2}$ to produce greyscale phase, $\Delta\phi$ (Lites and Chipman 1979). 
  
\section{Results}

\begin{figure}
\includegraphics[width=0.9\textwidth]{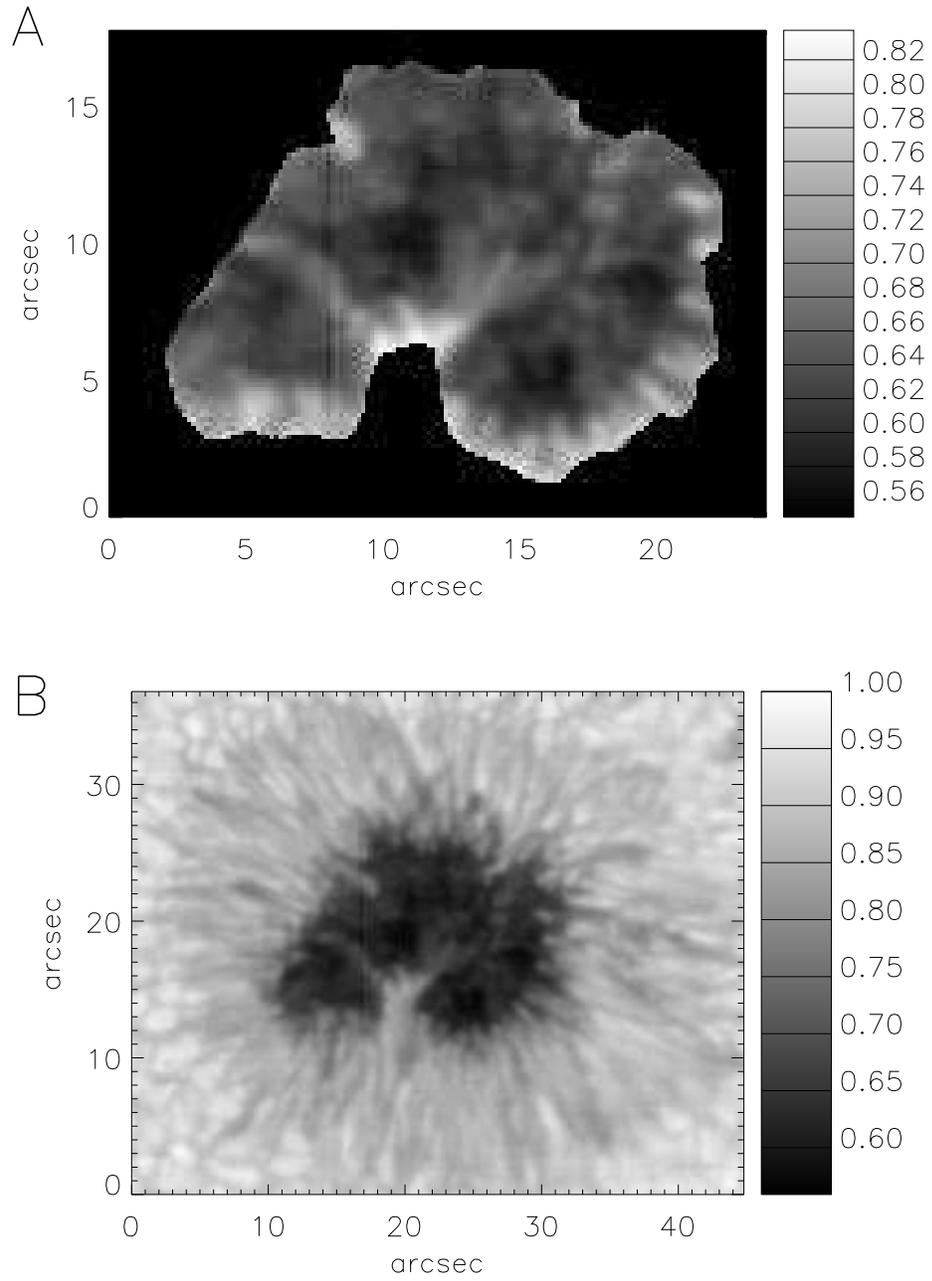}
\caption{Panel A represents the zoomed in part of the umbra, with the penumbral elements masked out, from the central spot of the active region visible in panel B. Both panels were taken in the NIR continuum. \label{umbra}}
\end{figure}

The target of our analysis, the umbra of the central spot of AR 10707 (Fig. \ref{umbra}A), was full of UDs. \par
\begin{figure}
\includegraphics[width=0.9\textwidth]{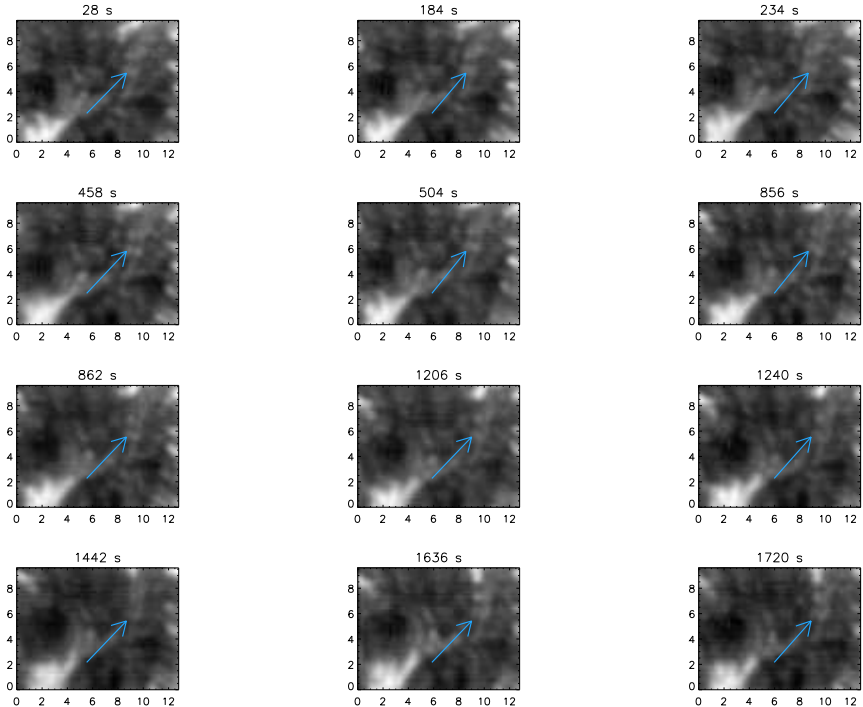}
\caption{An UD tracked through the time in the NIR dataset. Arrow at each frame marks an example UDs, in this instance  twin UDs. All selected frames have a contrast above 90\% of maximum contrast. The marked UDs stay in the same position for the duration of the time series.   \label{polozajj}}
\end{figure}

We observed and analysed $70$ easily resolved UDs. The UDs were long lived, $80$\% of them existed during our entire time series of $\sim 30$ min. During our time series, those $80$\% of UDs stayed in the same locations (Fig.\ref{polozajj}). \par 
The average size of UDs detected in this dataset is $0.86''$; double the Dawes' limit for the Dunn telescope in NIR continuum. Since we analyse objects that are close to each other, the Dawes' resolution limit is appropriate for our dataset. All the sizes of the values below Dawes' limit were ignored.\par 
We analysed the NIR continuum data set and G-band data set with wavelet analysis. The signal to noise ratio of our G-band data was poor since the seeing was not good during our observational run. Hence we used the G-band data only as test data for analysis methods. The data sets were $30$ minutes long, making the highest credible period in this dataset $12$ minutes. With the limitation to the period of 12 minutes we avoided the spurious oscillations that rise in the COI due to the finite nature of our time series. To avoid noise induced detections near  the Nyquist frequency, we ignored oscillations that did not contain at least 23 points in a single period.  Thus we analysed oscillations in the period range of $0.75$ to $12$ minutes. \par
\begin{figure}
\includegraphics[width=0.7\textwidth]{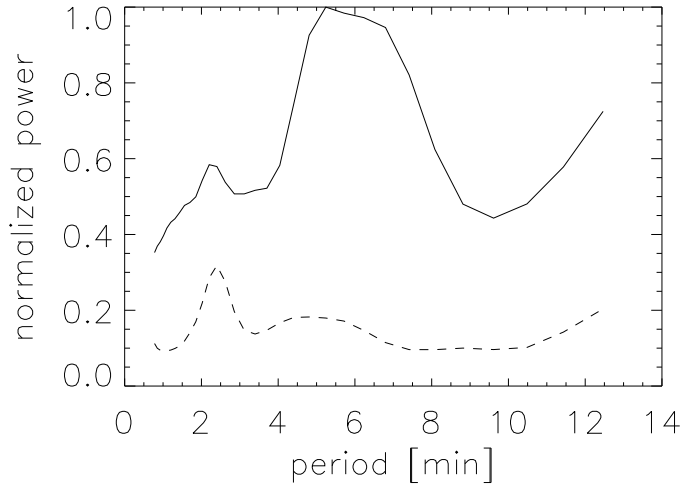}
\caption{Observed oscillatory power obtained with wavelet analysis, both curves normalised at the maximum of the NIR curve. The solid line represents the power observed in the NIR continuum, while the dashed line represents the power in G-band.\label{snaga}}
\end{figure}

 
The NIR continuum shows two strong peaks of oscillatory power at $3$ and $\sim 6.5$ minute (Fig.\ref{snaga}, solid line). The G-band data set (Fig.\ref{snaga}, dashed line) shows previously reported peaks at $3$ and $5$ minute (Balthasar and Wiehr, 1984; Lites, 1992; Tirospula et al., 2000) but no significant power at $\sim 6.5$ minute.\par 

 \end{multicols}
 
\begin{figure}
\includegraphics[width=0.9\textwidth]{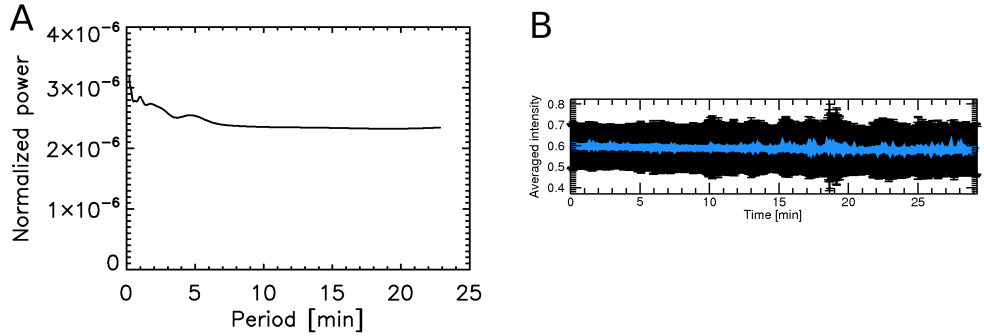}
\caption{Panel A presents the oscillatory power calculated for the intensity profiles of the individual UDs and averaged over the total number of UDs. The power profile shows peaks for $~3$ and $~5$ minutes oscillations. Panel B presents the averaged intensity variations (blue line) of the UDs, with the error bars. The intensity error for UDs intensities is $~20$\%. \label{snaga_ud}}
\end{figure}

\begin{multicols}{2}
The oscillations were separately calculated for UDs. The NAVE method provides, beside the tracking coordinates of the structures,  also the intensity changes of the structure through the time with the intensity error. We performed wavelet analysis on each of the UDs intensity curves to detect the oscillations emitted by UDs (Fig.\ref{snaga_ud}A). The power curve for the UDs shows a significantly smaller oscillatory power than the power emitted by the whole umbra. The maximum power peak for UDs emitted oscillations is at $\sim 1$ min, with a broad peak that covers periods from $\sim 2$ to $\sim 3$ minutes, finishing with the peak at $ \sim 5$ min period. The registered power is close to the noise level, since the intensity error is $\sim 20$\% (Fig.\ref{snaga_ud}B).
The oscillatory curve of NIR continuum shows the power peak at $3$ minutes period with another broad peak around $6$ minutes. The small peak at the $3$ minute  most probably represents oscillations emitted by UD. The other, more powerful oscillations registered in the other peak might, in part, originate from a different source. Considering the depth of the spectral line formation layer, we can speculate that the observed oscillations have a close connection to helioseismic oscillations. The spectrum of the helioseismic oscillations arises from modes with periods ranging from about $1.5$ minute to about $20$ minute (Gough and Toomre, 1991). The observed oscillatory peak may be a consequence of the plasma conditions and the shape of the cavity below the umbra. \par
\begin{figure}
\includegraphics[width=0.7\textwidth]{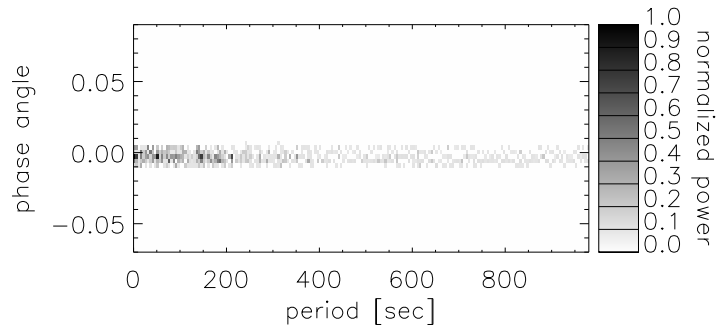}
\caption{Phase angle of the oscillations registered at NIR continuum. \label{faze}}
\end{figure}

 
Due to seeing conditions, the G-band data induce too large error into the phase difference calculations designed to determine if there is wave propagation. The umbra observed in our G-band data set did not show any distinguishing substructures or umbral flashes. We used a combination of the Fourier  and Hilbert transform to obtain the amplitudes and phases of the oscillations on both levels (Fig. \ref{faze}). \par 
At the NIR level, the oscillations with significant power have a phase of $0^o$. The range of the phase angle for all registered oscillations is from $-0.29^o$ to $0^o$. From this range, $89$\% of pixels in the umbra have $0^o$ phase, i.e. the registered waves are not propagating in any direction. The $11$\% of the pixels in umbra show signals of downward propagation with the phase angle less than $0.5^o$, indicating velocities $< 1$ m/s. \par 
\end{multicols}

\begin{figure}
\includegraphics[width=0.7\textwidth]{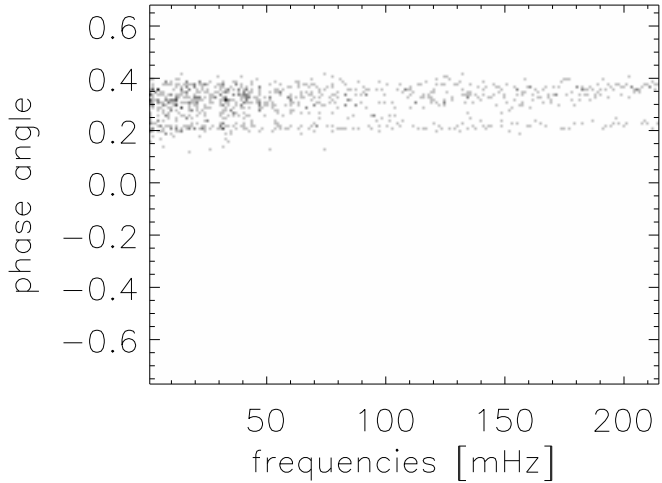}
\caption{Phase difference spectrum between oscillations registered at G-band level and the NIR level. A small phase angle can be caused with the excessive noise in G-band data and not the result of  real wave propagation. Positive phase difference indicates upward wave propagation. \label{faza}}
\end{figure}

\begin{multicols}{2}
We also calculated the phase difference spectra of the oscillations detected in both spectral bands to see is there any detectable wave propagation that might indicate a connection between the oscillations from both spectral bands. The calculations showed a very slight phase difference angle, less than $0.5^o$ (Fig. \ref{faza}). This small angle might be caused by the excessive noise in G-band dataset or by the small effects of dissipation through the atmosphere and not the true wave propagation. Hence we cannot state that there is a connection between oscillations detected in both spectral bands.\par 
These findings indicate that there is no observable wave propagation between the levels. However, due to the high noise levels in G-band data, this observation has to be reevaluated with another dataset. \par 

The oscillations observed in the NIR continuum dataset are distributed over the whole umbra. 
To ensure that the oscillations we analysed are real, we strengthened the restrictions for automated wavelet technique, as well as the procedures for the preparation of time sets. Use of de-stretching procedure reduces the possibility of intensity oscillations induced by shifting the object along the X-Y plane (where X and Y stand for the spatial coordinates). However, the induced intensity oscillations could not be 100\% removed, because the alignment and de-stretching were not performed at sub-pixel values. \par 

\end{multicols}

\begin{multicols}{2}

\begin{figure}
\includegraphics[width=0.7\textwidth]{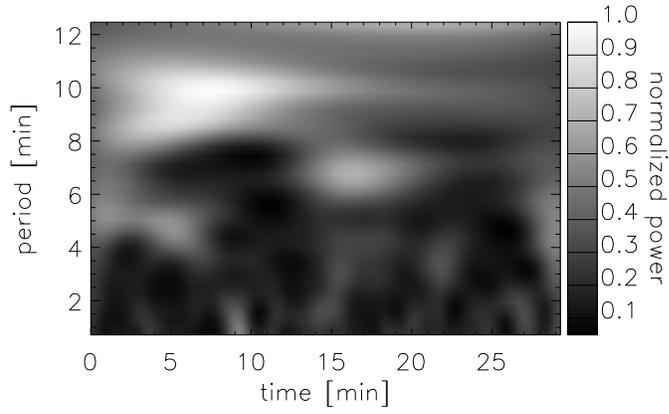}
\caption{Changes of the observed oscillatory power in the umbra over time. We integrated the power for each period individually over the whole surface of the umbra. \label{snagavr}}
\end{figure}

The umbra emits oscillations sporadically over the time (Fig. \ref{snagavr}). There are no constant oscillatory sources.  Hence we can state that over the time the oscillatory emission in the NIR continuum level is random for almost all periods. However, we have to limit this statement. Our time series is short and cannot cover the longer period oscillations for more than few periods;  hence we could not draw reliable conclusion about the behaviour of the longer period oscillations over the time.  \par 
We also checked the spatial distribution of the dominant oscillation frequency across the umbra (Fig. \ref{oscipro}). The umbra does not oscillate as a whole but in patches which are distributed across the umbra.  For each pixel in the umbra, we marked the period of the registered oscillations that carried the most power. That period was declared dominant for that analysed pixel. Each patch in the umbra emits oscillations over the whole period range we analysed, however the maximum power is contained within differed period oscillations for the neighbouring patches. We could not detect a connection of the period preference and the structures of the umbra. \par 
The patches are clearly defined and separated from each other with a sharp change in the oscillatory period. Some of the patches that oscillate with the low period oscillations have a gradual transition from one period oscillations to another within the patch itself, but still end abruptly at the edge. \par
Spatially these patches do not correspond to anything special inside the umbra, and cannot be connected with either UD or the spaces between. Some of the low period oscillations are connected with the observed UDs, however some of the UDs have short period oscillations. What causes the difference, we do not know. We hope that with the higher resolution, we will be able to resolve the cause of the difference in the patches.  \par

\section{Discussion and Conclusions}

We observed well-defined UDs in the NIR continuum level, $50$ km below $\tau_{500}=1$, and we detected oscillations in the umbra with power peaking at $3$ and $\sim 6.5 $ minutes.\par   
The UDs were distributed across the umbra and lasted longer than the duration of our dataset (Fig.\ref{polozajj}). We did not observe the appearance or disappearance of existing prominent UDs. Their average size is $0.8''$, and during our dataset they tended to stay at approximately the same location (Fig.\ref{polozajj}). 
Sch\"{u}ssler and V\"{o}gler (2006) model of magneto-convection in the umbra predicts that UDs are caused by rising plumes. This aspect was confirmed by the observational work of Rimmele (2008), Riethm\"{u}ller et al. (2008), and Ortiz et al. (2010). In Fig.5 of the work (Sch\"{u}ssler and V\"{o}gler, 2006), we can see the rising of a plume and the corresponding appearance of UDs in the brightness images. The lifetime of the average UD should be around 30 minutes of detectable brightness according to this model. However from the figure, it is clear that at lower levels in the umbra we would be able to see UDs for longer time. The example from Sch\"{u}ssler and V\"{o}gler, (2006)  demonstrates that plumes exist for almost $40$ minutes in the sub-photosphere. Moreover, Bharti et al. (2010) found that larger UDs live longer. We were only capable of resolving large UDs,  which were stable for the $30$ minute duration of our data sequence. Thus, we can speculate that the longevity of the observed UDs is a consequence of efficient convection at the NIR continuum formation level. Although this line of reasoning implies that the same would happen for smaller UD's, we could not observe them, hence we cannot broaden the statement to include the unobserved UDs. \par
Our result disagrees with Hamedivafa (2008), who found that the lifetime of UDs is between $7$ and $10$ minutes. However, Hamedivafa expresses doubt about his result, since possible effects were introduced by the automated analysis used for statistical analysis of the UDs. On the other hand, Watanabe et al. (2009) also used an automated detection code and found the mean lifetime to be $7.3$ minutes. \par 

The oscillations observed in the NIR continuum have power peaks at $3$ and $\sim 6.5$ minute period oscillations, while the testing of simultaneous observations in the G-band show umbral oscillations with power peaks near $3$ and  $5$ minutes. The observed $3$ minute oscillations are most probably connected with the oscillations of the UDs (Fig.\ref{snaga_ud}), while the broader peak around $6$ minute oscillations might originate from a different source.  \par
Watanabe et al. (2009) found in UDs intensity curves low frequency components close to $\sim 8$ minutes, while Sobotka and Puschmann (2009) reported that UDs substructures vary with the time scale of $\sim 3$ minutes.  Thus, we can speculate that observed power peaks in the intensity oscillations could be connected with same physical mechanism that caused the UDs and UD's structures to exist and change for the noted period in the spectral lines used in the cited  works, but at NIR continuum height we see those as a pure oscillations and not the variations of the UDs and UD's structures.\par 

Oscillations registered in the NIR continuum have small negative phase angle obtained by combining Fourier and Hilbert transform. Phase angles close to zero indicate evanescent waves. The phase angle close to zero is observed for the 89\% of powerful oscillations indicating that oscillations which contribute the most to the power peak around $6$ minute do not propagate upward.  The small negative phase angle is most probably associated with dissipation of evanescent waves. We can speculate that these oscillations are helioseismic oscillations, which buffet the photosphere and that these oscillations do not have a direct connection with the upward layers.  The typical helioseismic oscillations have periods ranging from $\sim1.5$ minute to $\sim 20$ minutes (Gough and Toomre, 1991). It is possible that the oscillations we observed are closely connected with the condition of the plasma directly below the umbra.  \par 
\end{multicols}

\begin{figure}
\includegraphics[width=0.7\textwidth]{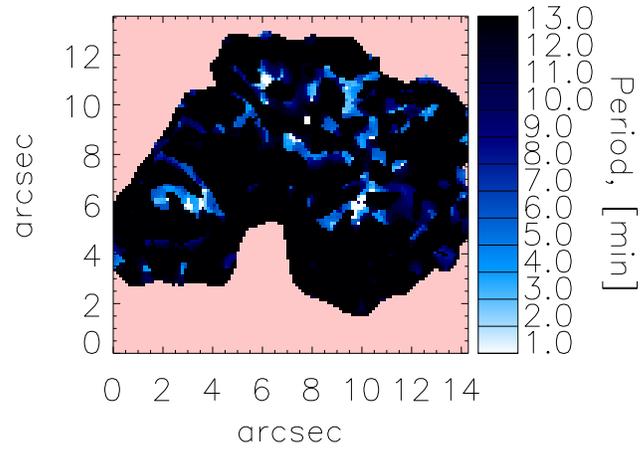}
\caption{Spatial distribution of the dominant oscillation periods across the umbra. \label{oscipro}}
\end{figure}

\begin{multicols}{2}

The phase angle for the most powerful oscillations is close to zero and the minimum oscillatory power is larger than the maximum oscillatory power registered in G-band. The phase spectra we used in this work weights phase angles with the power of the oscillatory signal in question, hence phase angles of the weak oscillations will be lost in our figure. We observed some waves with the positive phase angle, indicating upward propagation. However, the weaker power oscillations are closer to the sporadic noise detections of the oscillations and their reliability is questionable. Thus, we cannot state that there is a upward propagation of the detected oscillations.  \par
 
 The peaks at $3$ and $5$ minutes observed in G-band agree with the previous works (Balthasar and Wiehr, 1984; Lites, 1992; Tirospula et al., 2000), confirming the accuracy of our analysis methods. Our data set in G-band has poor S/N (signal to noise) ratio, thus cannot be used for establishing reliable connections between NIR dataset findings and the chromospheric findings.\par 
 
The power curve in Fig. \ref{snaga} for the NIR data set shows power for the $3$ minute oscillations as a small peak. This power peak is probably the contribution of the oscillations in the UDs. Although this might be taken as a confirmation of the origin of the oscillations from sub-photosperic levels (L\'{o}pez Aristre et al. 2001; Rouppe van der Voort et al. 2003; Bogdan and Judge, 2006), the low quality of the G-band data did not allow us to detect the vertical propagation of these oscillations and their connection with the upper atmospheric levels. Hence we cannot present firm proof that umbral oscillations originate on the NIR continuum level. \par 

The question remains about the other possible sources of the registered NIR oscillations. The instrumentation and the optical bench did not induce any oscillations. However, the dataset consist of the raw data, and the seeing influences differ from visible (G-band) to NIR spectral lines. Hence, it is necessary to adjust coalignment procedure to include the changes in the NIR, as well as in the visible part of spectra. Change of the interval used for the mean reference image for the coalignment of dataset caused a shift in the oscillatory power peak toward the longer periods. With an interval of 10 images (i.e. 20 sec), we got a strong power peak around the $10$ minute period.  With the interval of 20 images (i.e. 40 sec),  same peak shifted towards the longer period oscillations peaking at a $\sim 13$ minute period and with the interval of 30 images (i.e. 1 min), the peak shifted to $\sim 15$ minute period. These oscillations were the strongest around the high contrast areas (i.e. an edge of the umbra). The position of these oscillations and the changes caused by enlarging the image interval indicates that seeing distortions cause a shift in the images that might cause false oscillatory detections. There is a possibility that not even with the 30 images interval used for coalignment, we were not completely successful in removing all seeing induced oscillations. Thus, these findings have to be reevaluated using a dataset where the seeing influences are removed. \par

 Although the whole umbra oscillates, the dominant periods of the oscillations are not equally distributed over the umbra (Fig. \ref{oscipro}). There are distinct areas over the umbra that oscillate within the whole period range we analysed. This result agrees, in part, with Socas-Navarro et al. (2009), with a slight variation. Our patches have different dimensions and behave differently than the patches described in Socas-Navarro et al. (2009).  The difference in the oscillatory patches in this work and in the work by Socas-Navarro et al. (2009) might be caused by the different height of formation for the spectral lines used. NIR spectral band used in this work is located deep in the photosphere, and we do not have direct connection to the chromospheric observations. Thus, we cannot say that these patches are same as the ones observed in the work by Socas-Navarro et al. (2009). \par
 The location of the NIR spectral band might also lead to the speculation that the dominant oscillations with a $\sim 6.5$ minute period are helioseismic oscillations which do not propagate upward, while the oscillations detected in UD that have peaks close to $3$ and $5$ minutes might be the oscillations connected with the upper atmospheric layers. However, the high error in intensity curves of UD points to the necessity of checking this finding with some other dataset. \par

{\small {\it 

Thanks are due to the anonymous referee, which comments helped improve this work. AA thanks F. Davis DiPiazza for help with English. We gratefully acknowledge the support of NSF (AGS-0745744 and AGS-0847126), NASA (NNX08BA22G), and AFOSR (FA9550-09-1-0655). }}

\bibliography{simple}

\begin{thebibliography}{}
\bibitem[Andic {\it et al.}, 2010]{ja10}Andic,A., Goode,P.R., Chae,J., Cao,W., Ahn,K., Yurchyshyn,V., Abramenko,V. 2010, ApJL, 717, 79
\bibitem[Balthasar \& Wiehr, 1984]{balthasar84}Balthasar,H., Wiehr,E. 1984, SoPh, 94, 99
\bibitem[Bharti et al. 2010]{bharti10}Bharti, L., Beeck, B., Schssler, M., 2010, A\&A, 510, 12
\bibitem[Bloomfield et al. 2004]{shaun04}Bloomfield, D.S., McAteer, R.T.J., Lites, B.W., Judge, P.G., Mathoudakis, M., Keenan, F.P., ApJ, 617, 623
\bibitem[Bogdan \& Judge, 2006]{bogdan06}Bogdan, T.J., Judge, P.G. 2006, Phil. Trans. R. Soc. London A., 364, 313
\bibitem[Cao et al. 2005]{cao05}Cao, W., Xu, Y., Denker, C., \& Wang, H.\ 2005, Proc. SPIE, 5881, 245
\bibitem[Cao et al. 2006]{cao06}Cao, W., Hartkorn, K, Ma, J., Xu, Y., Spirock, T., Wang, H.,
\& Goode, P. R.\ 2006, SoPh, 238, 207
\bibitem[Chae \& Sakurai, 2008]{chae08}Chae, J., Sakurai, T., 2008, ApJ, 689, 593
\bibitem[Cheung et al. 2010]{cheung10}Cheung, M.C.M., Rempel, M., Title, A.M., Schssler, M., 2010, ApJ, 720, 233
\bibitem[Hamedivafa 2008]{hamed08}Hamedivafa, H., 2008, SoPh, 250, 17
\bibitem[Heineman et al. 2007]{heineman07}Heineman, T., Nordlund, ., Scharmer, G.B., Spruit, H.C., 2007, ApJ, 669, 1390
\bibitem[Graps 1995]{graps95} Graps, A., 1995, An introduction to wavelets, 1995, IEEE Computational Science and Engineering, vol.2, num.2
\bibitem[Gough \& Toomre 1991]{gough91}Gough, D., Toomre, J., 1991, A\&A Rev., 29, 627
\bibitem[Lites, B.W]{lites79}Lites, B.W, Chipman, E.G., 1979, ApJ, 231, 570
\bibitem[Lites, 1986]{lites86}Lites, B.W., 1986, ApJ, 301, 992
\bibitem[Lites, 1992]{lites92}Lites, B.W. 1992, In: Thomas J.H., Weiss N.O. (eds.) Sunspots:Theory and Observations, Kluwer, Dordecht, P.261
\bibitem[Lites, B.W]{lites98} Lites, B.W, Thomas, J.H, Bogdan, T.J., Cally, P.S. 1998,  ApJ, 497, 464
\bibitem[Lopez Aristre {\it et al.}, 2001]{lopez01}L\'{o}pez Aristre,A., Socas-Navarro,H., Molodij,G. 2001 ApJ, 552, 871
\bibitem[Ortiz et al. 2010]{ortiz10}Ortiz, A., Bellot Rubio, L.R., Rouppe van der Voort, L. H., M., 2010, ApJ, 713, 1282
\bibitem[Riethmuller et al. 2008]{riet08}Riethm\"{u}ller, T.L., Solanki, S.K., Lagg, A., 2008, A\&A, 492, 233
\bibitem[Rimmele, 2008]{rimmele08}Rimmele, T. 2008, ApJ, 672, 684
\bibitem[Rouppe van der Voort {\it et al.}, 2003]{rouppe03}Rouppe van der Voort,L.H.M., Rutten,P.J., Sutterlin,P., Sloover,P.J., Krijeg,J.M. 2003 A\&A, 403,277
\bibitem[Schussler \& Vogler, 2006]{schuss06}Sch\"{u}ssler, M., V\"{o}gler, A., 2006, ApJL, 641, 73
\bibitem[Socas-Navarro et al. 2009]{socas09}Socas-Navarro, H., McIntosh, S.W., Centeno, R., de Wijn, A.G., Lites, B.W., 2009, ApJ, 696, 1683
\bibitem[Soltau \& Wiehr, 2010]{soltau10}Soltau, D., Wiehr, E., 1984, A\&A, 141, 159
\bibitem[Starck et al. 1997] {starck97}Starck, J.L., Siebenmorgen, R., Gredel, R., 1997, AJ, 482, 1011
\bibitem[Stebbins \& Goode, 1987]{stebins87}Stebbins, R., Goode, P.R., 1987, SoPh, 110, 237
\bibitem[Torrence \& Compo, 1998]{torrence97}Torrence,C., Compo,G.P.: 1998, Bull. Amer. Meteor. Soc. , 79, 61
\bibitem[Tsiropula et al. 2000]{Tsiropoula00}Tsiropoula,G., Alissandrakis,C.E., Mein,P. 2000, A\&A, 355, 375
\bibitem[Vigouroux \& Delache 1993] {vigourox93}Vigouroux, A., Delache, Ph., A\&A 278, 607
\bibitem[Watanabe et al. 2009]{watanabe09}Watanabe, H., Kitai, R., Ichimoto, K., 2009, ApJ, 702, 1048
\bibitem[White and Cha, 1973]{white73}White, O.R., Cha, M.Y., 1973, SoPh, 31, 23
\bibitem[Zhugzhda {\it et al.} 1983]{zhugzhda83}Zhugzhda,Y.D., Locans,V., Staude,J. 1983, SoPh, 82, 369 
\end{thebibliography}

\end{multicols}
\end{document}